# Preserving Privacy and Sharing the Data in Distributed Environment using Cryptographic Technique on Perturbed data

P.Kamakshi , Dr.A.Vinaya Babu

**Abstract** -- The main objective of data mining is to extract previously unknown patterns from large collection of data. With the rapid growth in hardware, software and networking technology there is outstanding growth in the amount data collection. Organizations collect huge volumes of data from heterogeneous databases which also contain sensitive and private information about and individual .The data mining extracts novel patterns from such data which can be used in various domains for decision making .The problem with data mining output is that it also reveals some information, which are considered to be private and personal. Easy access to such personal data poses a threat to individual privacy. There has been growing concern about the chance of misusing personal information behind the scene without the knowledge of actual data owner. Privacy is becoming an increasingly important issue in many data mining applications in distributed environment. Privacy preserving data mining technique gives new direction to solve this problem. PPDM gives valid data mining results without learning the underlying data values .The benefits of data mining can be enjoyed, without compromising the privacy of concerned individuals. The original data is modified or a process is used in such a way that private data and private knowledge remain private even after the mining process. In this paper we have proposed a framework that allows systemic transformation of original data using randomized data perturbation technique and the modified data is then submitted as result of client's query through cryptographic approach. Using this approach we can achieve confidentiality at client as well as data owner sites. This model gives valid data mining results for analysis purpose but the actual or true data is not revealed.

———————————— ♦ ————————————

## 1 INTRODUCTION

Data mining is an emerging field which connects different major areas like databases, artificial intelligence and statistics. Data mining is a [15] powerful tool that can investigate and extract previously unknown patterns from large amounts of data. The process of data mining requires a large amount of data to be collected into a central site. In modern days organizations are extremely [9] [10] dependent on data mining in results to provide better service, achieving greater profit, and better decision-making. For these purposes organizations collect huge amount of data. For example, business organizations [22] collect data about the consumers for marketing purposes and improving business strategies, medical organizations collect medical records for better treatment and medical research. With the rapid advance of the Internet, networking, hardware and software technology there is remarkable growth in the amount of data that can be collected from different sites or organizations. Huge volumes of Data collected in this manner also include sensitive data about individuals. It is obvious that if a data mining algorithm is run against the union of different databases, the extracted knowledge not only consists of discovered patterns and correlations that are hidden in the data but it also reveals the information which is considered to be private. Privacy is an important issue in many data mining applications that deal with health care, security, financial and other types of sensitive
data. The actual anxiety of people is that their private information should not be misused behind the scenes without their knowledge. The real threat is that once information is unrestricted, it will be impractical to stop misuse. Privacy can for instance be threatened when data mining techniques uses the identifiers which themselves are not very sensitive ,but are used to connect personal identifiers such as addresses, names etc., with other more sensitive personal information .The simplest solution to this problem is to completely hide the sensitive data or not to include such sensitive data in the database.                                                                .

———————————————————————

P.Kamakshi ,Assistant Professor ,Department of Information Technology,Kakatiya Institute of Technology & science , Warangal,A.P.,India-506015

Dr. A.Vinaya Babu, Professor, Department of Computer Science and Engineering, Jawaharlal Nehru Technological University, Hyderabad,A.P.,India.

But this solution is not ideal and accurate because in many applications, like medicine research, DNA research etc. Different organizations or institutions wish to conduct a joint research on their databases because combining their data will definitely provide better results and mutual benefit to the organizations. In this scenario organizations want to share the data but neither of the institute or organizations want to disclose its database or private information about their clients due to privacy concern. In such a situation it is not only necessary to protect private and sensitive information but it is also essential to facilitate the use of database for investigation or for other purposes. Privacy preserving data mining [20] is a special data mining technique which has emerged to protect the privacy of sensitive data and also give valid data mining results. In this paper we propose a novel framework to preserve the privacy at both at customer end and data owner site. The original data is systematically transformed using randomized data perturbation privacy preserving data mining technique. This tailored data which maintains the characteristics and properties of original data is then submitted as result of customer query through cryptographic techniques.

## 2 PREVIOUS WORK

Recently the application of data mining is increased in various domains like business, academia, communication, bioinformatics and medicine .The data mining results not only gives the valuable information hidden in these databases, but sometimes also reveals private information about individuals. The difficulty is that data mining process extracts or evaluates the individual data which is considered as private  by means of linking different attributes. The true problem is not data mining, but the way data mining is done. PPDM is an emerging technique in data mining where privacy and data mining can coexist. It gives the summarized results without any loss of privacy through data mining process.
In general there are two main approaches in PPDM:
    i) Data transformation based
    ii) Cryptographic-based methods.
The data transformation based approach  modifies sensitive data in such a way that  it loses its sensitive meaning .In this process statistical properties of interest can be retained but  exact values cannot be determined during the mining process. Various data modification techniques are noise addition [1] [2] [3], data swapping [4], aggregation [5], suppression and signal transformation.
In Cryptographic techniques the data is encrypted with encryption methods and a set of protocols are used to allow the data mining operation. The set of protocols called as secured multiparty computation (SMC) is a computation process performed by group of parties where each party has in its control a part of the input data needed to perform the computation. In SMC the participating parties learns only the final result of the computation and no additional information is revealed during the computation. Perfect privacy in the SMC [6] [7] is achieved because any meaningful information is not released to any third party. The basic SMC PPDM techniques are secure sum, secure set union and secure size of set union.

### 2.1 Overview of Randomization Perturbation Technique

In randomization perturbation approach the privacy of the data can be protected by perturbing [13] sensitive data with randomization algorithms before releasing it to the data miner. The perturbed data version is then used to mine patterns and models. The algorithm is so chosen that combined properties of the data can be recovered with adequate accuracy while individual entries are considerably distorted. In this method privacy of confidential data [16] can be obtained by adding small noise component which is obtained from the probability distribution. In a set of data records denoted by $X = \{x_1 \ldots x_N\}$. For record $x_i \in X$, we add a noise component which is drawn from the probability distribution. Commonly used distributions are the uniform distribution over an interval $[-\alpha, \alpha]$ and Gaussian distribution with mean $\mu = 0$ and standard deviation $\sigma$. These noise components are drawn independently, and are denoted $y_1 \ldots y_N$. Thus, the new set of distorted records are denoted by $x_1 + y_1 \ldots x_N + y_N$. We denote this new set of records by $z_1 \ldots z_N$. In general, it is assumed that the variance of the added noise is large enough, so that the original record values cannot be easily guessed from the distorted data. One key advantage of the randomization method is that it is relatively simple, and does not require knowledge of the distribution of other records in the data.

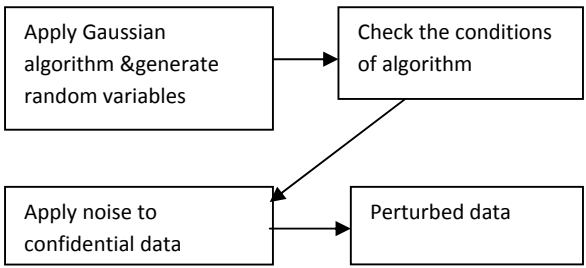

**F**ig. 1. Block diagram for implementing perturbation technique

### 2.2 Cryptographic Technique

Cryptography, the science of communication and computing in the presence of a malicious [27] adversary extends from the traditional tasks of encryption and authentication to protocols for securely distributing computations among a group of mutually distrusting parties. In an ideal situation, in addition to the original parties there is also a third party called "trusted party" who does not deviate from the activities prescribed for him. All parties send their inputs to the trusted party, who then computes the function and sends the appropriate results to the other parties. The protocol that is run in order to compute the function does not leak any unnecessary information. Sometimes there are limited leaks of information that are not dangerous. This process requires high level of trust.

Cryptographic community therefore aims at designing protocols that do not reveal [28] any information except for their designated output, without the involvement of trusted party. Secure multiparty computation is performed by multiple parties where each party has in its possession a part of the input needed to perform the computation.

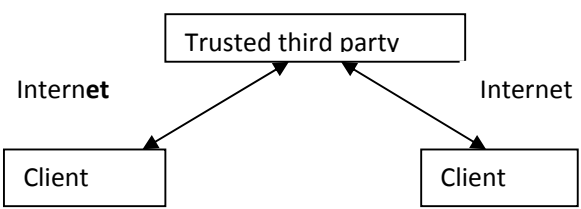

Fig. 2. System using Semi trusted third party

However, at the end of the computation, the parties should only have learned the result of the computation. Secure Multi-Party Computation (SMC) allows parties with similar background to compute upon their private data, minimizing the threat of disclosure. The security requirements are that nothing is learned from the protocol other than the output and that the output is distributed according to the prescribed functionality. In secure multiparty computation a set of N parties with private inputs $x_1,\ldots,x_n$ on a network can compute a joint function of their inputs. This joint computation should have the property that the parties learn the correct output $y = f(x_1,\ldots, x_n)$ and nothing else.

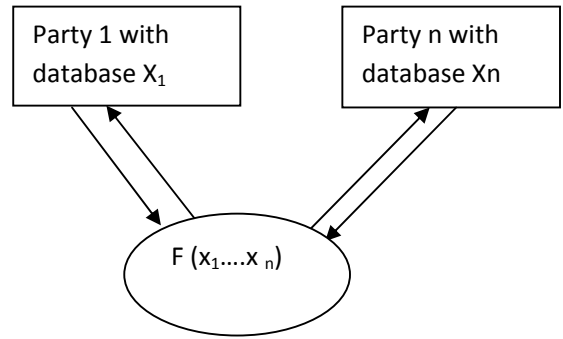

Fig. 3. Secure multiparty computation system

## 3 OUR APPROACH

In this novel framework the total process is divided into three components the customer, mediator and a group of service data providers. At any time the role of service data provider and customer can be interchanged. The role of mediator is purely passive; it keeps the record of no. of data providers and transfers the query from customer to data providers. Initially there is no communication between customer and data provider. When the client sends a query, the mediator send the information to all the data holders and through exchange of acknowledgements, the mediator establishes the connection with data providers.

## 4 ALGORITHM

Step 1: Client sends a query to mediator to obtain the number of actual data providers.

Step 2: Mediator sends the query to all the data providers.

Step 3: The data providers which satisfies the client requirements sends the acknowledgement to mediator.

Step 4: Mediator keeps the information of N as a record indicating the total no. of relevant data provider and sends the value N to the client.

Step 5: Each data provider sends to the mediator N sets where each set consists of randomly generated sets of keys. The mediator sends the N sets of keys to client. The client selects one key from each set of keys.

Step 6: The client encrypts its own public key with the selected key from each set of keys. Each set of keys are then sent back to the mediator.

Step 7: The mediator sends back the set of keys to the data providers.

Step 8: The data providers perturb the data and encrypts each table entry using all the keys in each set and sends the encrypted data back to the customer through mediator.

Step 9: Client decrypts the data using its private key and obtains the required information in a consolidated form.

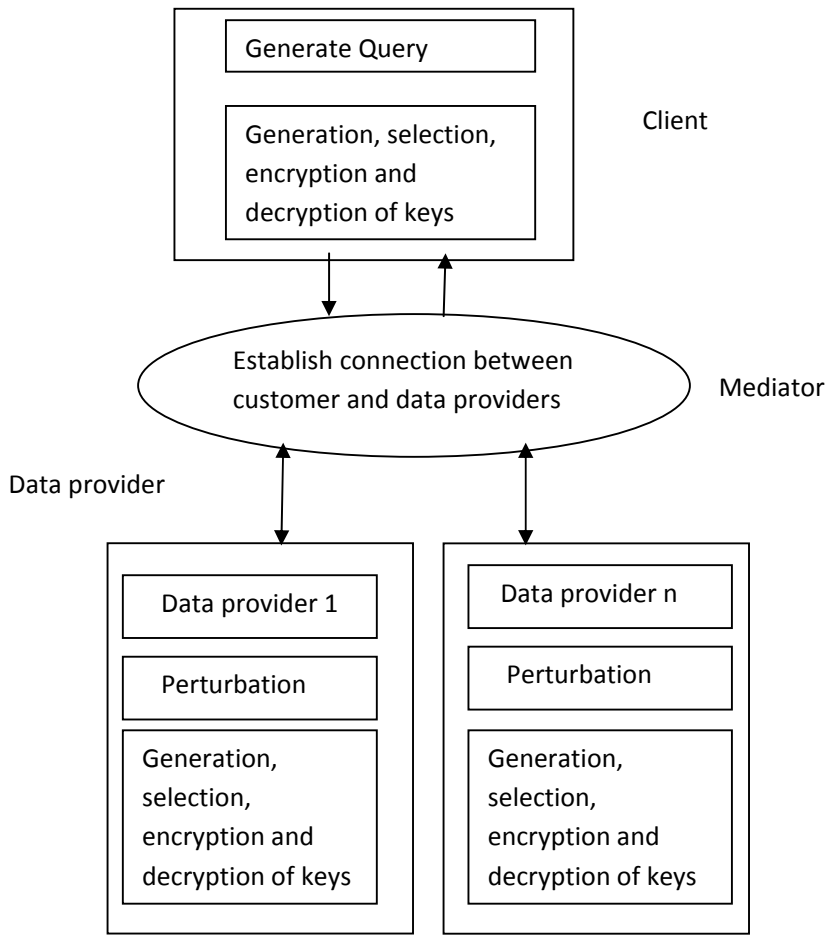

Fig. 5. Framework for sharing the data using encryption and perturbation technique

## 5 DISCUSSION

During the entire process the privacy is maintained at both the customer and service provider's location. The privacy is achieved because the customer who is query generator doesn't know the exact details of data providers who actually contributed the results except the value N ie. total no. of data owners. Similarly the data owner don't know which key out of given set of keys have been selected by the customer for encryption purpose. The result obtained by the customer is also in the perturbed form in which small amount of noise is added to sensitive data such that the properties and the meaning of the original data is not changed but privacy is maintained.

Fig. 6. A sample database of a Hospital

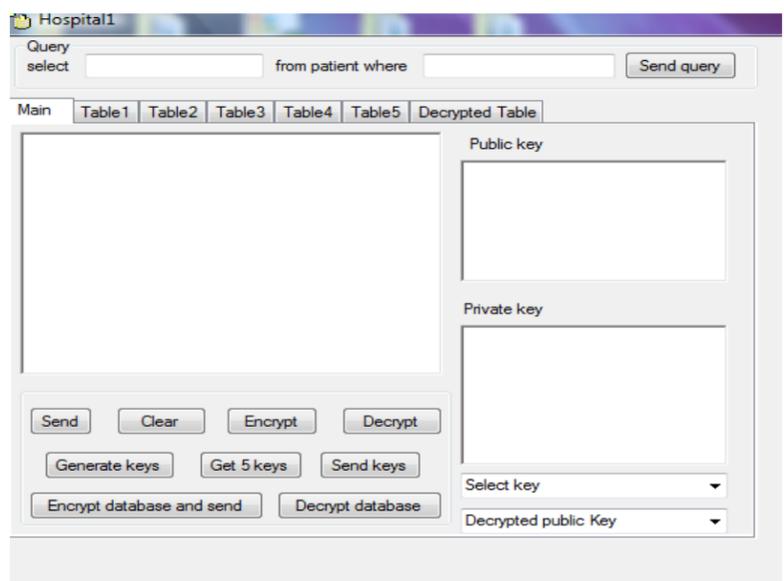

Fig. 7. A sample database of other hospital

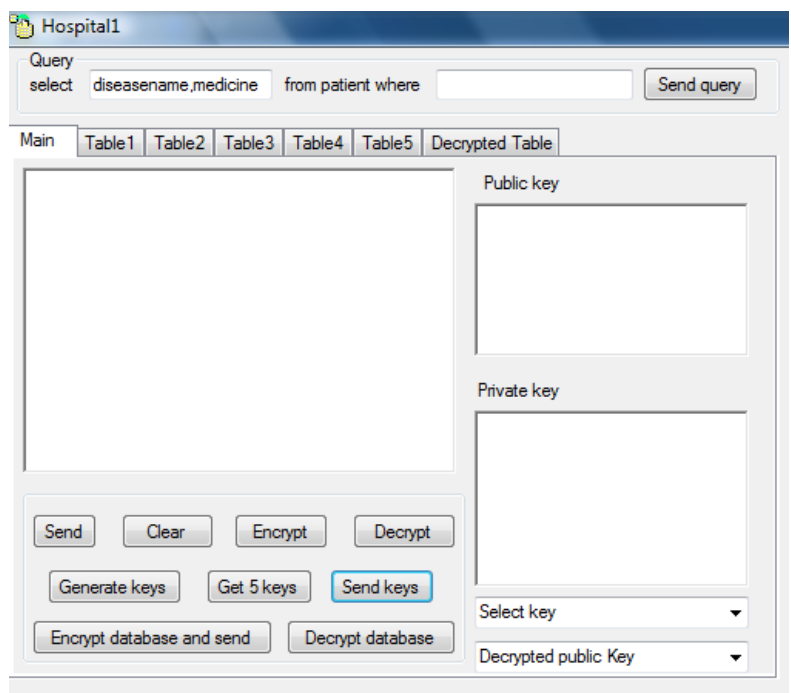

Fig. 8. All the hospitals maintains similar format of database

Fig 9. Query submitted by one of client

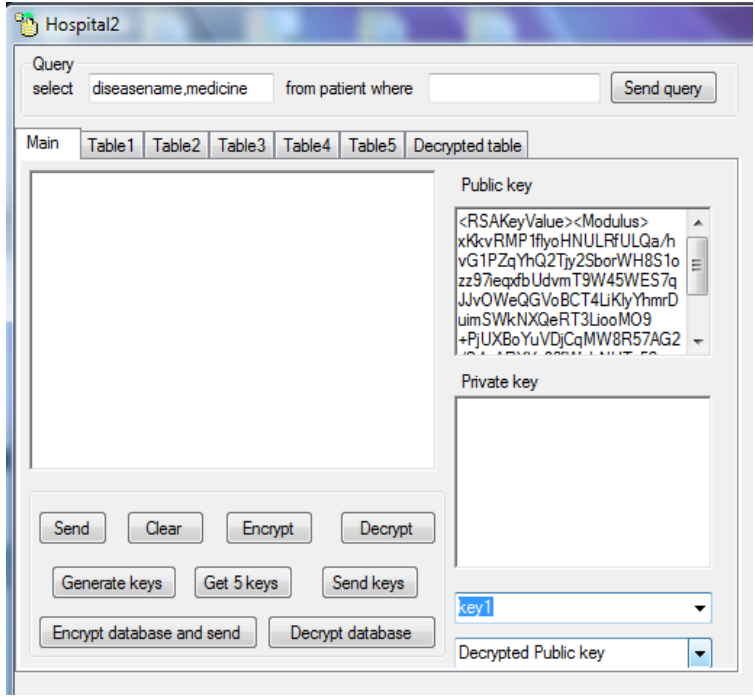

Fig. 10. Exchange of keys, encryption & decryption

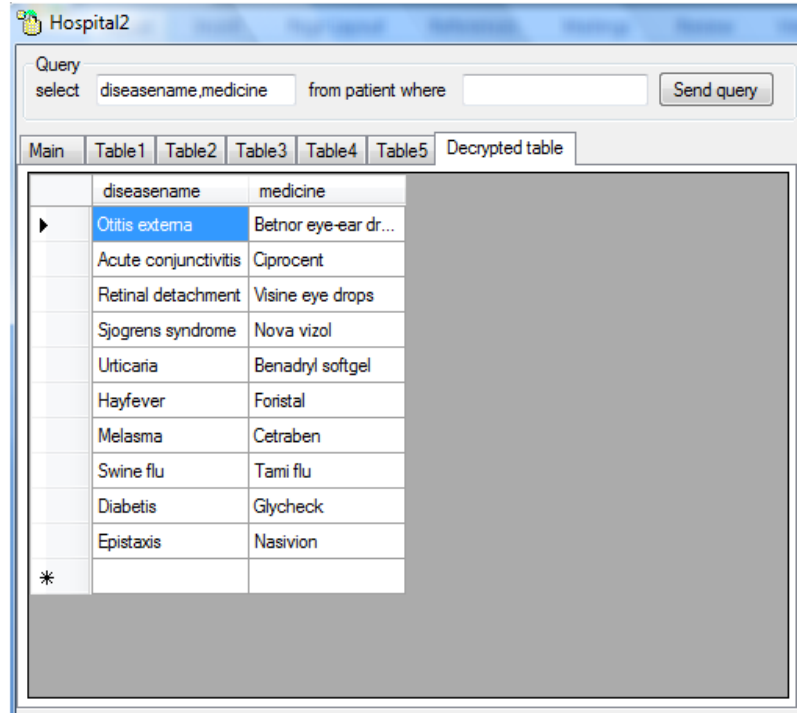

Fig. 13. The actual query result. The privacy is obtained because the key selected by client is not known to data owners & the client doesn't know the different actual data sources who contributed data for the client's query.

## 6 CONCLUSION

The ever increasing ability to identify and collect large amounts of data, analyzing the data using data mining process and decision on the results gives prospective benefits to organizations .But, such repositories also contains private and sensitive information and releasing the personal information can cause significant damage to data owner. Hence there is increased need to discover and distribute the databases, without compromising the privacy of the individual's data. In this paper we presented a novel approach in which data perturbation technique used to modify the original data at data owner server and then cryptographic technique is used to submit the result of customer query .This technique is applicable in distributed environment where each data owner has its own data and also wants to share the data with other data owners, but at the same time want to preserve the privacy of sensitive data in the data records. The confidentiality is guaranteed among the distributed sites because the details of data owner are hidden from client , and data owner is completely unaware of the selection of keys by client during cryptographic process. We obtain enhanced privacy because the result obtained is in perturbed form, so the privacy of original data is retained giving valid data mining results.

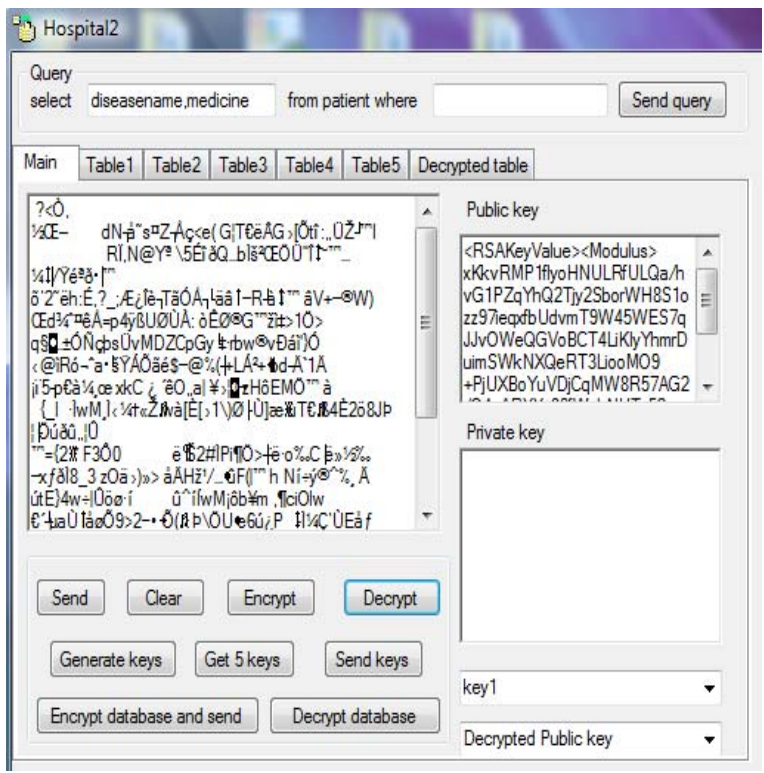

Fig 11. Decryption of database with different keys

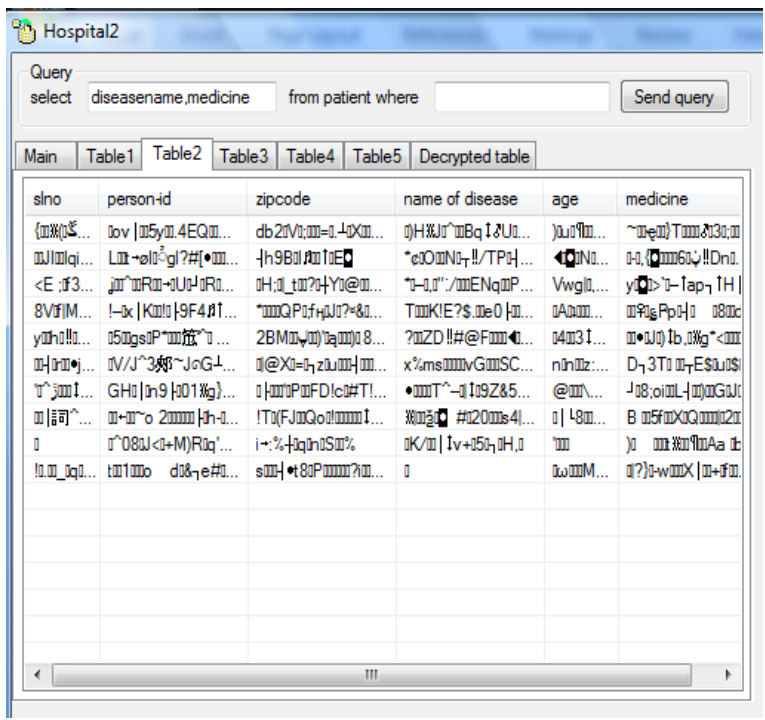

Fig. 12. Decrypted database table